\documentstyle[12pt]{article}
\newcommand{\h}{\hspace{.5cm}}
\newenvironment{destaque}{\begin{quotation}\small\em}{\end{quotation}}
\date{}
\title{Boundary Conditions as Mass Generation Mechanism for Complex Scalar 
Fields}
\author{{\bf Jos\'e Alexandre Nogueira}\\
{\bf Denimar Possa}\\
{\it Departamento de F\'{\i}sica, Centro de Ci\^encias Exatas,}\\
{\it Universidade Federal do Esp\'{\i}rito Santo,}\\
{\it 29.060-900 - Vit\'oria-ES - Brazil,}\\
{\it E-mail: } nogueira@cce.ufes.br}
\begin{document}
\maketitle
\begin{abstract}
\begin{destaque}
We consider the effects of homogeneous Dirichlet's boundary conditions in 
the scalar electrodynamics with self-interaction. We have found for a critical scale of the compactification length that symmetry is restored and 
scalar field develops mass and vector field does not.

\end{destaque}
\end{abstract}
\section{Introduction}
\h The spontaneous symmetry breaking  plays a essential role in the Higgs mechanism for the mass generation of the fundamental particles. In the standard model the spontaneous symmetry breaking is induced by the Higgs potential through an imaginary mass term inserted by hand. The arbitrariness of the Higgs potential is one of the weakness of the Higgs mechanism, since many of the physical parameters depend on the precise form of the Higgs potential \cite{ref1, ref2}.

In an alternative approach of Coleman-Weinberg \cite{ref3}, the spontaneous symmetry breaking is induced by 1-loop radiative corrections, rather than being inserted by hand. 
Furthermore, in the Coleman-Weinberg approach one start with a theory massless in the tree approximation and then the theory becomes massive due to radiative corrections. 
Nevertheless, in the Coleman-Weinberg approach both the scalar and the vector particles develop masses proportional to the vacuum expectation value. So, massless scalar electrodynamics does not remain massless, but nor does it remain electrodynamics.

The idea of boundary conditions as mass generation mechanism is not new. 
Ford and Yoshimura \cite{ref4} have introduced the idea that the existence of a non-trivial space-time topology can lead to the generation of mass and many others works have been publishing \cite{ref5, ref6, ref7, ref8, ref9}. In recent works we have stressed that there is a critical scale of the compactification length where a massless theory undergoes a transition to one massive in order $\hbar^{0}$  \cite{ref10, ref11}.

In this work we study the effects of the boundary conditions on the scalar and vector fields of the massless scalar electrodynamics which satisfy homogenous Dirichlet conditions on two infinite parallel plane surfaces separated by some small distance $a$. As will be seen subsequently the boundary conditions may restore the symmetry and the scalar field may acquire mass. Thus, massless scalar electrodynamics turns out mass scalar electrodynamics.

The outline of the paper is as follows. In section 2 we calculate the effective potential for the scalar electrodynamics at 1-loop. In section 3 we consider the fields satisfying Dirichlet's boundary conditions on two infinite parallel plane surfaces separated by some distance $a$. 
We derive the effective potential at 1-loop level as sums of modified Bessel functions, and we take $a$ very small to expand each term of the sums in power of $a$. After throwing away higher-order terms, we evaluate the vacuum expectation value and the renormalized mass. 
In section 4 we point out our conclusions and some speculations.

\section{Effective potential for the scalar electrodynamics}

\h Let us consider the theory of a massless, quartically 
self-interacting complex scalar field $\phi(x)$ minimally coupled to the electrodynamics field.
The lagrangian density for this theory is
\begin{equation}
{\cal L=}- \frac{1}{4}F_{\mu\nu}F^{\mu\nu} +
 D_{\mu }\phi^{*}D^{\mu }\phi - 
\frac{\lambda }{6}\left(\phi^{*}\phi\right)^{2},
\end{equation}
where $D_{\mu}$ is the covariant derivate, given by
\begin{equation}
D_{\mu} = \partial _{\mu } + ieA_{\mu},
\end{equation}
necessary to keep the lagrangian invariant under transformation
 of the U(1) group and it is answerable for the minimally coupled.

The contribution to effective potential can be somewhat simplified 
writing the complex field in terms of two real fields $\phi_{1}$ and 
$\phi_{2}$. Putting
\begin{equation}\phi = \frac{1}{\sqrt{2}}\left( \phi_{1} + 
i\phi_{2} \right),
\end{equation}
the lagrangian (1) becomes
\begin{equation}
{\cal L=}- \frac{1}{4}F_{\mu\nu}F^{\mu\nu} +
 \frac{1}{2} \left( \partial _{\mu } \phi_{1} 
- ieA_{\mu} \phi_{2} \right)^{2} +
 \frac{1}{2} \left( \partial _{\mu } \phi_{2} + 
ieA_{\mu} \phi_{1} \right)^{2} -
\frac{\lambda }{4!}\left(\phi_{1}^{4} + \phi_{2}^{4} + 
2\phi_{1}^{2} \phi_{2}^{2} \right).
\end{equation}

The one-loop effective potential is given by\footnote{We keep 
$\hbar$ to mark the quantum corrections, but we set $\hbar = c = 1$ everywhere else.}
$$
V_{ef}\left( \phi _{c}\right) = \frac{\lambda }{4!}\phi_{c}^{4} 
+ \frac{3}{2}\frac{\hbar }{\Omega }\ln \det \left[ k_{A}^{2} + 
e^{2}\phi_{c}^{2} \right] + 
$$
\begin{equation}
+ \frac{1}{2}\frac{\hbar }{\Omega }
\ln \det \left[ k^{2} + \frac{\lambda}{2}\phi_{c}^{2} \right] 
+ \frac{1}{2}\frac{\hbar }{\Omega }
\ln \det \left[ k^{2} + \frac{\lambda}{6}\phi_{c}^{2} \right] ,
\end{equation}
where
\begin{equation}
\phi_{c}^{2} = \phi_{1c}^{2} + \phi_{2c}^{2} ,
\end{equation}
and $\Omega$ is the four dimensional space-time volume.

The second term of the above $V_{ef}\left( \phi _{c}\right)$ 
corresponds to the term $\frac{1}{2}e^{2}\phi_{c}^{2}A_{\mu}^{2}$ 
of the potential and the last term corresponds to the 
$\frac{\lambda }{12}\phi_{1}^{2}\phi_{2}^{2}$, that is, 
they are contributions of graphics of external $\phi_{1}$ legs 
and $A_{\mu}$ loop (second term) and $\phi_{2}$ loop (last term).
The third term of $V_{ef}\left( \phi _{c}\right)$  corresponds 
to graphics of external $\phi_{1}$ legs and $\phi_{1}$ loop. 
The first term is the classical potential. Since there are 
three independent components of the vector field, we have the 
factor 3 in the second term.

Now, using the proceeding of zeta function regularization 
\cite{ref12, ref13, ref14,  ref15, ref16} we obtain
$$
V_{ef}\left( \phi _{c}\right) = \frac{\lambda}{4!}\phi_{c}^{4} + 
\frac{3\hbar}{64\pi^{2}}e^{4}\phi_{c}^{4}
\left[ \ln\left(\frac{e^{2} \phi _{c}^{2}}{\mu ^{2}}\right) - 
\frac{3}{2} \right] +
$$
\begin{equation}
+ \frac{\hbar}{256\pi^{2}}\lambda^{2}\phi_{c}^{4}
\left[ \ln\left(\frac{\lambda \phi _{c}^{2}}{2\mu^{2}}\right) - 
\frac{3}{2} \right]
+ \frac{\hbar}{2304\pi^{2}}\lambda^{2}\phi_{c}^{4}
\left[ \ln\left(\frac{\lambda \phi _{c}^{2}}{6\mu^{2}}\right) - 
\frac{3}{2} \right].
\end{equation}\\
The above equation is the effective potential in 1-loop for scalar electrodynamics with quartic self-coupling required for 
renormalizability on a free space-time (without boundary).

 As it is well known, even though for tree approximation the minimum occurs at $\left\langle \phi \right\rangle =0$, the radiative correction of the second term in the above effective potential will lead to non-trivial vacuum, rather, vacuum will become degenerate, if $\lambda$ is of order $e^{4}$.
Hence, the spontaneous symmetry breaking induced by radiative corrections leads both scalar and vector particles to dynamically acquire masses whose values are proportional to the vacuum expectation value \cite{ref3, ref14, ref16, ref17}.

\section{Mass generation for the complex scalar field}

\h Let us consider the scalar $\phi$ and vector gauge $A_{\mu}$ fields satisfying Dirichlet's boundary condition on two infinite parallel plane surfaces separated by some small distance $a$. Now, the effective potential, Eq.(5), is written as

$$
V_{ef}\left( \phi _{c}\right) = \frac{\lambda }{4!}\phi_{c}^{4} 
+ \frac{3}{2}\frac{\hbar }{\Omega }\ln \det \left[ k_{A}^{2} + 
\frac{n_{3}^{2}\pi^{2}}{a^{2}} + e^{2}\phi_{c}^{2} \right] + 
$$
\begin{equation}
+  \frac{1}{2}\frac{\hbar }{\Omega }
\ln \det \left[ k^{2} + \frac{n_{1}^{2}\pi^{2}}{a^{2}} + \frac{\lambda}{2}\phi_{c}^{2} \right] 
+ \frac{1}{2}\frac{\hbar }{\Omega }
\ln \det \left[ k^{2} + \frac{n_{2}^{2}\pi^{2}}{a^{2}} + \frac{\lambda}{6}\phi_{c}^{2} \right].
\end{equation}

Using proceeding of zeta function regularization we obtain \cite{ref10}
$$
V_{ef}\left( \phi _{c}\right) = \frac{2\alpha^{2}}{4!}\phi_{c}^{4} + 
\frac{3\hbar}{32\pi^{2}}e^{4}\phi_{c}^{4}
\left[ \ln\left(\frac{e^{2} \phi _{c}^{2}}{\mu ^{2}}\right) - 
\frac{3}{2} \right]  - \frac{\hbar}{4\pi ^{2}}e^{4}\phi _{c}^{4}\sum_{n_{3}=1}^{\infty }
\frac{K_{2}\left( 2n_{3}ea\phi _{c}\right)}
{\left( n_{3}ea\phi_{c}\right) ^{2}} +
$$
$$
+  \frac{\hbar}{24\pi a}\alpha ^{3}\phi_{c}^{3} \left[ 1 + 
\frac{1}{3 \sqrt{3}} \right] +
\frac{\hbar }{64\pi^{2}}\alpha^{4}\phi _{c}^{4}
\left[ \ln\left(\frac{\alpha^{2} \phi _{c}^{2}}{\mu ^{2}}\right) +
\frac{1}{9} \ln\left(\frac{\alpha^{2} \phi _{c}^{2}}
{3\mu ^{2}}\right) - \frac{5}{3} \right] +
$$
\begin{equation}
- \frac{\hbar}{8\pi ^{2}}\alpha ^{4}\phi _{c}^{4}
\sum_{n_{1}=1}^{\infty }\frac{K_{2}
\left( 2n_{1}\alpha a\phi _{c}\right)}
{\left( n_{1}\alpha a\phi_{c}\right) ^{2}} - 
 \frac{\hbar}{72\pi ^{2}}\alpha ^{4}\phi _{c}^{4}
\sum_{n_{2}=1}^{\infty }\frac{K_{2}
\left( 2n_{2}\sqrt{3}\alpha a\phi _{c}\right)}
{\left( n_{2}\sqrt{3}\alpha a\phi_{c}\right) ^{2}},
\end{equation}
where we define $\alpha^{2}=\frac{\lambda}{2}$ for the sake of simplicity.
The term of order $\alpha^{4}$ may be dropped against the first term of order $\alpha^{2}$. Therefore,
$$
V_{ef}\left( \phi _{c}\right) = \frac{2\alpha^{2}}{4!}\phi_{c}^{4}
+  \frac{\hbar}{24\pi a}\alpha ^{3}\phi_{c}^{3} \left[ 1 + 
\frac{1}{3 \sqrt{3}} \right] +
$$
$$
+  \frac{3\hbar}{32\pi^{2}}e^{4}\phi_{c}^{4}
\left[ \ln\left(\frac{e^{2} \phi _{c}^{2}}{\mu ^{2}}\right) - \frac{3}{2} \right]
 - \frac{\hbar}{4\pi ^{2}}e^{4}\phi _{c}^{4}\sum_{n_{3}=1}^{\infty }
\frac{K_{2}\left( 2n_{3}ea\phi _{c}\right)}
{\left( n_{3}ea\phi_{c}\right)^{2}} +
$$
\begin{equation}
- \frac{\hbar}{8\pi ^{2}}\alpha ^{4}\phi _{c}^{4}
\sum_{n_{1}=1}^{\infty }\frac{K_{2}
\left( 2n_{1}\alpha a\phi _{c}\right)}
{\left( n_{1}\alpha a\phi_{c}\right) ^{2}} - 
\frac{\hbar}{72\pi ^{2}}\alpha ^{4}\phi _{c}^{4}
\sum_{n_{2}=1}^{\infty }\frac{K_{2}
\left( 2n_{2}\sqrt{3}\alpha a\phi _{c}\right)}
{\left( n_{2}\sqrt{3}\alpha a\phi_{c}\right) ^{2}}.
\end{equation}

Let us consider the small $a$ and $\alpha$ of order $e$ case.
 Since $\phi$ is of order $\alpha^{-1}$, $\alpha a \phi_{c} << 1$ and $ea \phi_{c} << 1$. 
Consequently, we expand the last three terms on the right-hand side of Eq.(10) \cite{ref10}.
 Hence, up to higher-terms, the effective potential becomes
\begin{equation}
V_{ef}\left( \phi _{c}\right) = \frac{2\alpha^{2}}{4!}\phi_{c}^{4} + 
\frac{\hbar}{48 a^{2}} e^{2}\phi_{c}^{2}
+ \frac{\hbar}{72 a^{2}} \alpha^{2}\phi_{c}^{2} - \frac{\hbar\pi^{2}}{360 a^{4}}.
\end{equation}

The minimum occurs at $\phi _{c}=\left\langle \phi \right\rangle $, 
where 
\begin{equation}
\left. \frac{dV_{ef}}{d\phi _{c}}\right| _{\phi _{c}=
\left\langle \phi \right\rangle }=0.
\end{equation}

Differentiating Eq.(11), we have
\begin{equation}
\left\langle \phi \right\rangle \left[\frac{\alpha^{2}}{3}
\left\langle \phi \right\rangle^{2} + 
\frac{\hbar  e^{2}}{24 a^{2}} + \frac{\hbar \alpha^{2}}{36 a^{2}} \right] = 0.
\end{equation}

There will be non-trivial solution of Eq.(13) if the sum of the terms between brackets vanishes. However, there is not order of $a$ (with respect to the orders of $e$) which permits a non-trivial solution. Therefore the minimum of $V_{ef}$ is satisfied for
\begin{equation}
\left\langle \phi \right\rangle = 0.
\end{equation}

The result (14) shows that the vacuum is non-degenerate, and, therefore spontaneous symmetry breaking cannot occur. Actually, the boundary conditions allow just one constant vacuum solution, $\left\langle \phi \right\rangle = 0$. If the result had been 
$\left\langle \phi \right\rangle = constant \neq 0$, the effective potential would not have been used and a solution
 $\left\langle \phi \right\rangle = \phi_{0}(x)$ would have been expected because the boundary conditions break the translational invariance (that is the case when $\alpha$ is of order $e^{2}$). Since symmetry breaking does not take place the vector field does not develop mass through the Higgs mechanism.

Although the effective potential given by the Eq.(11) is finite, that 
is not a final result because the coupling constant in it is an arbitrary 
parameter. Therefore, we must fit it to the renormalized coupling constant 
using the renormalization condition
\begin{equation}
\left. \frac{d^{4}V_{ef}}{d\phi _{c}^{4}}\right| _{\phi _{c}=M}
=\lambda _{R},
\end{equation}
where $M$ is a arbitrary floating renormalization point, to get $\lambda = \lambda_{R}$.

The renormalized mass of the scalar field is given by
\begin{equation}
\left. \frac{d^{2}V_{ef}}{d\phi _{cl}^{2}}\right| 
_{\phi_{c}=\left\langle\phi\right\rangle = 0 }=
\frac{\hbar \lambda_{R}}{36 a^{2}} + \frac{\hbar e^{2}}{24a^{2}} = m_{\phi}^{2},
\end{equation}
which may be non-zero in order $\hbar^{0}$ depending on the order of $a$ with respect to the order of $e$. 

Eq.(16) shows if $a$ is of order $e$ then the scalar field mass will be of order $\hbar^{0}$. This result is in agreement with our initial assumption. 
If $a$ is of lower order than $e$ ($a > e$) then the mass 
will lie far outside the expected range of validity of our 
approximation. It follows that there is a typical length scale given by the compactification length $a$ of the theory. Within this length scale, the massless scalar 
electrodynamics theory with self-interaction becomes massive 
due to boundary conditions effects.

It is worth noting the fact that if $a$ is small enough 
the theory will become one massive in order $(\hbar^{0})$. 
This is because $e$ and $a$ are independent parameters, 
so the mass term found can be of order zero-loop, even though it is a one-loop result (it is from radiative 
corrections). Since $m_{\phi}^{2}$ is of order $\hbar^{0}$, it cannot be subtracted out by mass counter-term of order $\hbar$.

We stress the fact that if $a$ is small enough, spontaneous symmetry breaking will not take place and the effect of the boundary conditions will be of restoring the symmetry of the theory. Consequently, the gauge vector field $A_{\mu}$ will not develop mass through the Higgs mechanism. 

\subsection{Boundary conditions only on the gauge field}
\h Let us only consider the gauge vector field $A_{\mu}$ satisfying the boundary conditions. Now, the effective potential can be obtained from Eq.(10) taking $a$ goes to infinity in the second and last two terms. Thus, we get
\begin{equation}
V_{ef}\left( \phi _{c}\right) = \frac{2 \alpha^{2}}{4!}\phi_{c}^{4} + 
\frac{3\hbar}{32\pi^{2}}e^{4}\phi_{c}^{4}
\left[ \ln\left(\frac{e^{2} \phi _{c}^{2}}{\mu ^{2}}\right) - 
\frac{3}{2} \right] - 
\frac{\hbar}{4\pi ^{2}}e^{4}\phi _{c}^{4}\sum_{n=1}^{\infty }
\frac{K_{2}\left( 2nea\phi _{c}\right)}
{\left( nea\phi_{c}\right) ^{2}}.
\end{equation}

Since the boundary conditions are only imposed on the vector field, 
the vacuum expectation value of $\phi$ field may develop any constant value and the effective potential can be used to determine it.

Now, let us consider the small $a$ and $\alpha$ of order $e$ case. We expand the last term on the right-hand side of Eq.(17). Up to higher-terms we find
\begin{equation}
V_{ef}\left( \phi _{c}\right) = \frac{2\alpha^{2}}{4!}\phi_{c}^{4} + 
\frac{\hbar}{48 a^{2}} e^{2}\phi_{c}^{2}
 - \frac{\hbar\pi^{2}}{720 a^{4}}.
\end{equation}

The minimum of the above effective potential is satisfied for 
\begin{equation}
\left\langle \phi \right\rangle = 0.
\end{equation}
The renormalized mass of the scalar field is given by
\begin{equation}
\left. \frac{d^{2}V_{ef}}{d\phi _{cl}^{2}}\right| 
_{\phi_{c}=\left\langle\phi\right\rangle = 0 }=
\frac{\hbar e^{2}}{24a^{2}}=m_{\phi}^{2}.
\end{equation}

Since symmetry breaking does not take place, the Eq.(20) shows if $a$ is of order $e$ then the scalar field will develop mass and the vector field will not.

If $ea\phi_{c} > 1$ then spontaneous symmetry breaking will take place and the effect of the boundary conditions ought to be a 
displacement of the vacuum expectation value of $\phi$ field 
from one have been found by S. Coleman and E. Weinberg.

The last term of Eq.(18) depends on the global space-time structure through 
$\frac{1}{a^{4}}$. It is nothing but the Casimir energy (vacuum energy density) of the electromagnetic field.

\section{Conclusion}
\h In this paper we have studied the scalar electrodynamics confined between two parallel plane surfaces separated by small distance $a$. We have showed that in the small $a$ and $\lambda$ of order $e^{2}$ case, spontaneous symmetry breaking does not take place. Because, when $a$ is not small enough spontaneous symmetry breaking is induced by radiative corrections, i. e., by Coleman-Weinberg mechanism, we can infer there is a critical length scale of the finite region where the symmetry is restored. We have found that if $a$ is of order $e$, the massless scalar field will acquire mass but vector field will not, because spontaneous symmetry breaking does not take place. Since $a$ goes like $\frac{1}{m_{\phi}}$, it is comparable with the Compton wavelength of the scalar particle. 

We are aware that our work appears as simplified useful exercise for investigating the mechanism through which boundary conditions generate mass for the elementary particle. Although, we feel that our work may be applied to study superconductivity and others subjects, too. We would like to warn that we have not account the gauge invariance of our results \cite{ref2, ref18}

It is worth noting the fact our approach for scalar electrodynamics does lead to the mass generation for scalar particles but not for photons, like we expect.\\

{\bf Acknowledgments}

\h  This work was supported in part by the National Agency for Research (CAPES)(Brazil).\\

\end{document}